\documentclass[]{llncs}

\usepackage{lineno,hyperref}
\usepackage{color}
\usepackage{bm}
\usepackage{amsmath}
\usepackage{amssymb}
\usepackage{graphicx}
\usepackage{subfigure}
\usepackage{subfigmat}
\usepackage{multirow}
\usepackage{algorithm}
\usepackage{algorithmic}
\usepackage{footnpag}

\begin{document}

\title{Forecasting satellite trajectories by interpolating hybrid orbit propagators}

\author{Iv\'an P\'erez\inst{1}, Montserrat San-Mart\'in\inst{2}, Rosario L\'opez\inst{3}, Eliseo P. Vergara\inst{1}, Alexander Wittig\inst{4}  \and Juan F\'elix San-Juan\inst{1}}
\authorrunning{Iv\'an P\'erez et al.}

\institute{Scientific Computing Group (GRUCACI), University of La Rioja, \\26006 Logro\~no, Spain
\and
Scientific Computing Group (GRUCACI), University of Granada,\\  52005 Melilla, Spain
\and
Scientific Computing Group (GRUCACI), Center for Biomedical Research of La Rioja,  26006 Logro\~no, Spain
\and
Advanced Concepts Team, European Space Agency, 2200 AG Noordwijk, The Netherlands
}

\titlerunning{Interpolating}  

\maketitle

\begin{abstract}
A hybrid orbit propagator based on the analytical integration of the \textit{Kepler problem} is designed to determine the future position and velocity of any orbiter, usually an artificial satellite or space debris fragment, in two steps: an initial approximation generated by means of an integration method, followed by a forecast of its error, determined by a prediction technique that models and reproduces the missing dynamics. In this study we analyze the effect of slightly changing the initial conditions for which a hybrid propagator was developed. We explore the possibility of generating a new hybrid propagator from others previously developed for nearby initial conditions. We find that the interpolation of the parameters of the prediction technique, which in this case is an additive Holt-Winters method, yields similarly accurate results to a non-interpolated hybrid propagator when modeling the $J_2$ effect in the \textit{main problem} propagation.
\end{abstract}

\section{Introduction}

The propagation of perturbed orbits is a well-known problem which implies having to tackle a set of three second-order or six first-order differential equations, so as to determine the position and velocity of an orbiter at a given final time $t_f$ from its situation at an initial instant $t_1$.

As these equations are not directly integrable, there are three well-established techniques aimed at providing a solution to the problem. Each of these methods can be characterized in terms of the formulation of the equation of motion, the integration method used to obtain the solution to this equation, which can be numerical or analytical, and, finally, the perturbation model taken into account.

\textit{General perturbation theories} apply perturbation methods to the determination of an analytical solution. Such solution, which is an explicit function of time and some physical constants, allows for a fast determination of the coordinates at $t_f$. In addition, being an analytical expression, it embeds the dynamics of the problem. Nevertheless, in order to avoid extreme complexity, the analytical solution is usually a low-order approximation in which only the most relevant forces are considered.

\textit{Special perturbation theories}, in contrast, perform a numerical integration of the problem. They have the advantage of allowing for the consideration of any effect into the model, even the complex ones, thus leading to highly accurate solutions. Nonetheless, the disadvantage lies in the necessity to take small integration steps, which implies long computational time.

\textit{Semianalytical techniques} take advantage of both theories. They allow for the consideration of complex perturbing effects into the model, which is simplified by means of analytical methods so as to remove the short-period component. Consequently, the new equations of motion can be numerically integrated through longer steps, resulting in reduced computational time.

More recently, the \textit{hybrid propagation methodology} has been presented. It is based on the combination of any of the aforementioned integration methods and a forecasting technique. The former generates an initial solution, which is approximate because of the assumed simplifications and inaccuracies in the perturbation models. The latter makes use of forecasting techniques, based on either statistical time series models \cite{san2012gru_sarimahop,san2016gru_tshop} or machine learning methods \cite{per2013gru_nnhop}, in order to provide, once adjusted with a set of real observations that include the dynamics neglected in the initial approximation, a prediction of its error. The sum of this error prediction and the initial solution generates the final result.

The forecasting component of a hybrid propagator needs a set of control data, deduced from precise observations or accurately computed coordinates, so that the statistical or machine learning technique can model dynamics not present in the first stage of the method.

Nevertheless, a grid of hybrid propagators for a set of relatively close initial conditions can be constructed, in such a way that hybrid propagators for intermediate cases can be directly deduced from the grid, with no need for control data. By doing so, the study of initial conditions in the surroundings of an orbiter can be easily handled with no need to recompute the parameters of the forecasting component of the hybrid method.

In this paper we will consider the so-called \textit{main problem} of the artificial satellite theory, that is, the \textit{Kepler problem} only perturbed by the flattening of the Earth. We will create a hybrid propagator, composed of a general perturbation theory derived from the \textit{Kepler problem} plus an additive Holt-Winters method modeling the $J_2$ effect, for a certain orbiter. In order to handle both eccentricity and inclination small variations, we will construct a grid of hybrid propagators around the studied satellite. After that, we will prove that the forecasting component of the hybrid propagator, when eccentricity and/or inclination slightly vary, can be directly derived from the grid by simply interpolating the parameters of the Holt-Winters method.

The outline of the paper is divided into seven sections. Section 2 presents the principles of the hybrid propagation methodology, whereas Section 3 focuses on the use of an exponential smoothing technique, the Holt-Winters method, as the forecasting stage of hybrid propagators. The described concepts are applied to the creation of a hybrid propagator for a certain satellite in Section 4. With the aim of studying its surroundings, a grid of initial conditions, together with its corresponding hybrid propagators, is created around the studied satellite in Section 5. Section 6 illustrates how to develop new hybrid propagators for nearby initial conditions through the interpolation of parameters from other propagators in the grid. Finally, Section 7 summarizes the conclusions of the study and future lines of research.

\section{Hybrid propagation methodology}

The hybrid propagation methodology is aimed at estimating the position and velocity of an orbiter, which can be an artificial satellite or a fragment of space debris, at a given final time $t_f$, $\hat{\bm{x}}_f$, starting from the known position and velocity at an initial instant $t_1$, $\bm{x}_1$. It is worth noting that any set of canonical or non-canonical variables can be used for this purpose.

In a first stage, an integration method $\mathcal{I}$ is used to calculate an initial approximation of $\hat{\bm{x}}_f$:

\begin{equation}\label{solana}
\bm{x}_f^{\mathcal{I}} = \mathcal{I}(t_f,\bm{x}_1).
\end{equation}

The integration method is applied to a mathematical model that not always describes the physical phenomena exactly. Moreover, when the general perturbation theory or semianalytical techniques are used, only the most important forces and low-order approximations are usually considered; otherwise cumbersome expressions would be obtained. Due to all these facts, $\bm{x}_f^{\mathcal{I}}$ is an initial approximation that needs to be complemented in a second stage in order to obtain $\hat{\bm{x}}_f$.

The information that this second stage needs to model and reproduce, that is, the missing dynamics, has to be deduced from a \textit{control interval} $[t_1, t_T]$, with $t_T < t_f$. Throughout this interval both the initial approximation $\bm{x}_i^{\mathcal{I}}$ and the exact position and velocity $\bm{x}_i$ are assumed to be known, for example by means of precise observations or intensive and accurate numerical propagations. Therefore, the error due to the missing dynamics for any instant in the control interval can be expressed as

\begin{equation}
\mathcal{\bm\varepsilon}_i = \bm{x}_i - \bm{x}_i^{\mathcal{I}}, \label{error} 
\end{equation}

\noindent
and the time series of the errors of each of the six variables during the control interval, which we will call \textit{control data}, can be constructed as $\mathcal{\bm\varepsilon}_1,\ldots,\mathcal{\bm\varepsilon}_T$.

The processing of this time series, by means of either statistical techniques or machine learning methods, allows for the modeling of its behavior and, more importantly, its prediction at any time outside the control interval. Therefore, an estimation of the error at the final instant $t_f$, $\hat{\mathcal{\bm\varepsilon}}_f$, can be determined, and thus the desired value of $\hat{\bm{x}}_f$ can be calculated as

\begin{equation} \label{forecast}
\hat{\bm{x}}_f= \bm{x}_f^{\mathcal{I}} + \hat{\mathcal{\bm\varepsilon}}_f.
\end{equation}

\section{Exponential smoothing method for time series forecasting}

Exponential smoothing methods consider a time series $\varepsilon_t$ as the combination of three components: the trend $\mu_t$, or secular variation, the seasonal component $S_t$, or periodic oscillation, and the irregular or non-predictable component $\nu_t$. In the case of an additive composition, $\varepsilon_t$ can be expressed as

\begin{equation}
\varepsilon_t=\mu_t + S_t + \nu_t.
\end{equation}

In particular, the Holt-Winters method \cite{win1960_forecexpma} considers a linear trend with level $A$ and slope $B$:

\begin{equation}
\mu_t=A+B t.
\end{equation}

According to this method, and taking into account that $\nu_t$ cannot be predicted, the next value of a time series can be estimated from past values as

\begin{equation} \label{SE1} 
\hat{\varepsilon}_{t}=A_{t-1}+ B_{t-1}+S_{t-s},
\end{equation}

\noindent
where $s$ is the period of the seasonal component, and $A$, $B$, and $S$ can be determined from previous values according to the following recurrences

\begin{eqnarray} 
	A_t & = & \alpha (\varepsilon_t-S_{t-s}) + (1-\alpha) (A_{t-1}+B_{t-1}), \nonumber \\[.5ex]
	B_t & = & \beta(A_t-A_{t-1}) + (1-\beta) B_{t-1}, \label{SE2} \\[.5ex]
	S_t & = & \gamma(\varepsilon_t-A_{t}) + (1-\gamma) S_{t-s}, \nonumber
\end{eqnarray}

\noindent
in which $\alpha$, $\beta$, and $\gamma$ are three smoothing parameters with values in the interval $[0,1]$.

\begin{algorithm}
\begin{algorithmic}[1]
\REQUIRE $s$, $c$, $h$, and $\{\varepsilon_t\}_{t=1}^{T}$
\ENSURE $\hat{\varepsilon}_{T+h|T}$
\STATE Estimate the values of $A_0, B_0,S_{-s+1},\ldots, S_{-1},S_0$
\FOR {$t=1;\,t\leq T;\,t=t+1$}
\STATE $A_t = \alpha (\varepsilon_t-S_{t-s}) + (1-\alpha)(A_{t-1}+ B_{t-1})$
\STATE $B_t = \beta(A_t-A_{t-1}) + (1-\beta) B_{t-1}$ 
\STATE $S_t = \gamma(\varepsilon_t-A_{t}) + (1-\gamma) S_{t-s}$
\STATE $\hat{\varepsilon}_{t} = A_{t-1}+ B_{t-1} + S_{t-s}$
\ENDFOR
\STATE Select \texttt{error$\_$measure} $\in$ \{MSE, MAE, MAPE\} and express it as a function of the smoothing parameters
\STATE Obtain the smoothing parameters that minimize \texttt{error$\_$measure} using the L-BFGS-B method
\STATE Calculate $A_T, B_T, S_{T-s+1},\ldots, S_{T-1}, S_T$ for the optimum smoothing parameters
\STATE  $\hat{\varepsilon}_{T+h|T} = A_T + h B_T + S_{T-s+1+h\,\mathrm{mod}\, s}$
\RETURN  $\hat{\varepsilon}_{T+h|T}$
\end{algorithmic}
\caption{Holt-Winters}\label{algHW}
\end{algorithm}

Algorithm~\ref{algHW} shows how to apply the Holt-Winters method to the prediction of future time series values. The inputs to the algorithm are the amount of data per revolution, $s$, the number of revolutions in the control interval, $c$, the number of time steps after the control interval for which the time series value has to be predicted, $h$, and the control data, $\{\varepsilon_t\}_{t=1}^{T}$, with $T=s \times c$. The output is $\hat{\varepsilon}_{T+h|T}$, that is, the forecast of the time series at the final instant $t_f=t_{T+h}$, based on the last control data, $\varepsilon_T$.

The algorithm starts by estimating the initial parameters $A_0$, $B_0$, $S_{-s+1},\ldots$, $S_{-1}$, and $S_0$, which is accomplished through a classical additive decomposition into trend and seasonal variation over the three first revolutions. A linear regression over the trend provides the initial level $A_0$ and slope $B_0$, whereas the seasonal component yields the values of $S_{-s+1},\ldots$, $S_{-1}$, and $S_0$.

Then, an iterative process takes place by applying Eqs.~\eqref{SE1} and \eqref{SE2} to the control interval (lines 2--7). As a result, the expressions of the parameters $A_t$, $B_t$, $S_t$, and the single-step time series prediction $\hat{\varepsilon}_{t}$ are obtained as functions of the smoothing parameters $\alpha$, $\beta$, and $\gamma$.

After that, an error measure is selected among mean square error, MSE, mean absolute error, MAE, and mean absolute percentage error, MAPE.

The selected error measure applied to the control interval yields an expression which is a function of the smoothing parameters. Then, an optimization method is necessary to determine the values of the smoothing parameters that minimize this error measure. The limited memory algorithm L-BFGS-B \cite{byr1995_limmemopt}, which is a variation of the BFGS method \cite{sha1970_quasinewt}, allows to impose restrictions on the smoothing parameters, and hence is the algorithm that has been used.

Once the optimal smoothing parameters have been found, the time series parameters $A_T$, $B_T$, $S_{T-s+1},\ldots,S_{T-1},S_T$ are determined for the last period of the control interval, from which the forecasted time series value at the final instant, that is, $h$ epochs ahead, $\hat{\varepsilon}_f=\hat{\varepsilon}_{T+h|T}$, can be calculated (line 11).

\section{Application of the hybrid propagation methodology}

In this section, the described hybrid methodology is applied to the propagation of an orbit with the following initial conditions: semi-major axis $a=7228$ km, eccentricity $e=0.06$, and inclination $i=49^{\circ}$. The first stage of the method is an analytical theory derived from the \textit{Kepler problem}, that is, considering no perturbations at all, whereas the second part is an additive Holt-Winters method, designed to model the perturbation caused by the flattening of the Earth, which corresponds to the $J_2$ term in its gravitational potential. Therefore, the complete hybrid propagator is adapted to the \textit{main problem} of the artificial satellite theory; consequently, its results will be compared with those obtained from a highly accurate numerical integration of the \textit{main problem} by means of a high-order Runge-Kutta method.

The solution to the \textit{Kepler problem} provided by the analytical expression in the first stage of the hybrid propagator is characterized by constant values in all the classical orbital elements, except in the mean anomaly, whose values evolve following the orbiter angular position. In contrast, when the $J_2$ effect is considered, no orbital element remains constant, so that, in general, secular, short-period, and long-period effects can be found in the evolution of orbital elements. The goal of the Holt-Winters method in the second stage of the hybrid propagator is the modeling and reproduction of such dynamics. The difference between the initial Kepler solution and the desired \textit{main problem} solution translates into a position error of about 14500 km after 20 days of propagation, which represents approximately the distance between the apogee and perigee of the orbit.

The hybrid methodology can be applied to any set of variables, although Delaunay variables $(l, g, h, L, G, H)$ will be used in this case. The first step consists in preparing the control data, which is composed of two time series: the initial approximations generated by the analytical expression derived from the \textit{Kepler problem}, and the accurate solutions calculated by means of a high-order Runge-Kutta method. The last time series could be substituted for a set of precise observations in case they were available. The subtraction of both data sets yields the time series of the error, which contains the dynamics missing from the initial approximation. It is worth noting that the control data set should be large enough so as to include any pattern to be modeled by the second stage of the method. In this case, a control interval of ten revolutions has been chosen, which represents a time span of nearly 17 hours, taking into account that the aforementioned orbital elements correspond to an orbital period of 101.926 minutes. The sampling rate for the time series has been taken equal to 12 samples per orbiter revolution, which corresponds to a sampling period of $101.926/12=8.494$ minutes.

Before processing data, angular variables are homogenized to the interval $(-\pi,\pi]$ by adding or subtracting complete spins to values outside this interval. An univariate Holt-Winters model is considered for the time series of the error of each Delaunay variable, except for $\varepsilon_t^H$, which is 0 in this case, which means that the analytical approximation is perfect for this variable, and hence there is no need to complement it in the second stage of the hybrid propagator.

Then, a preliminary analysis of the five remaining time series is performed through the study of their sequence graphics, periodograms, and autocorrelation functions (ACF). This analysis reveals the existence of three main seasonal components, with periods a third, a half, and one Keplerian period, that is, 33.976, 50.964, and 101.926 minutes, although the last one is the most remarkable and includes the others.

Next, Algorithm~\ref{algHW} is applied, selecting MSE as the error measure needed to determine the optimal values for the smoothing parameters $\alpha$, $\beta$, and 
$\gamma$.

Once the five Holt-Winters models corresponding to Delaunay variables $l$, $g$, $h$, $L$, and $G$ have been created, they are integrated into the hybrid propagator so as to evaluate its accuracy through the comparison with a precise numerical propagation by means of a high-order Runge-Kutta method.

Table~\ref{tdist_S} compares the position error, after different propagation spans, between the analytical approximation, which only considers the \textit{Kepler problem}, and the hybrid propagation, which models the \textit{main problem}. As can be seen, the latter presents reduced errors, even after 30 days of propagation, which implies that the forecasting part of the hybrid method has been able to model most of the $J_2$ effect. 

\begin{table}
\caption{Distance error (km) after propagating the studied satellite.} \label{tdist_S}
\centering 
\begin{tabular}{ccc}
\hline\noalign{\smallskip}
\multirow{2}{*}{Propagation span} & Analytic method & Hybrid method \\
 & (Kepler) & (Kepler + $J_2$) \\
\noalign{\smallskip}
\hline
\noalign{\smallskip}
$\phantom{l}1$ day &	$\phantom{1}1197.10$ & $\phantom{1}0.45$ \\[.5ex]
$\phantom{3}2$ days &	$\phantom{1}2379.94$ & $\phantom{1}0.83$ \\[.5ex]
$\phantom{3}7$ days &	$\phantom{1}7900.47$ & $\phantom{1}3.63$ \\[.5ex]
30 days &	$14504.69$ & $13.73$ \\
\hline
\end{tabular}
\end{table}

\section{Creation of a grid from control data}

\begin{figure}[!!htp]
\centering
\includegraphics[scale=.6]{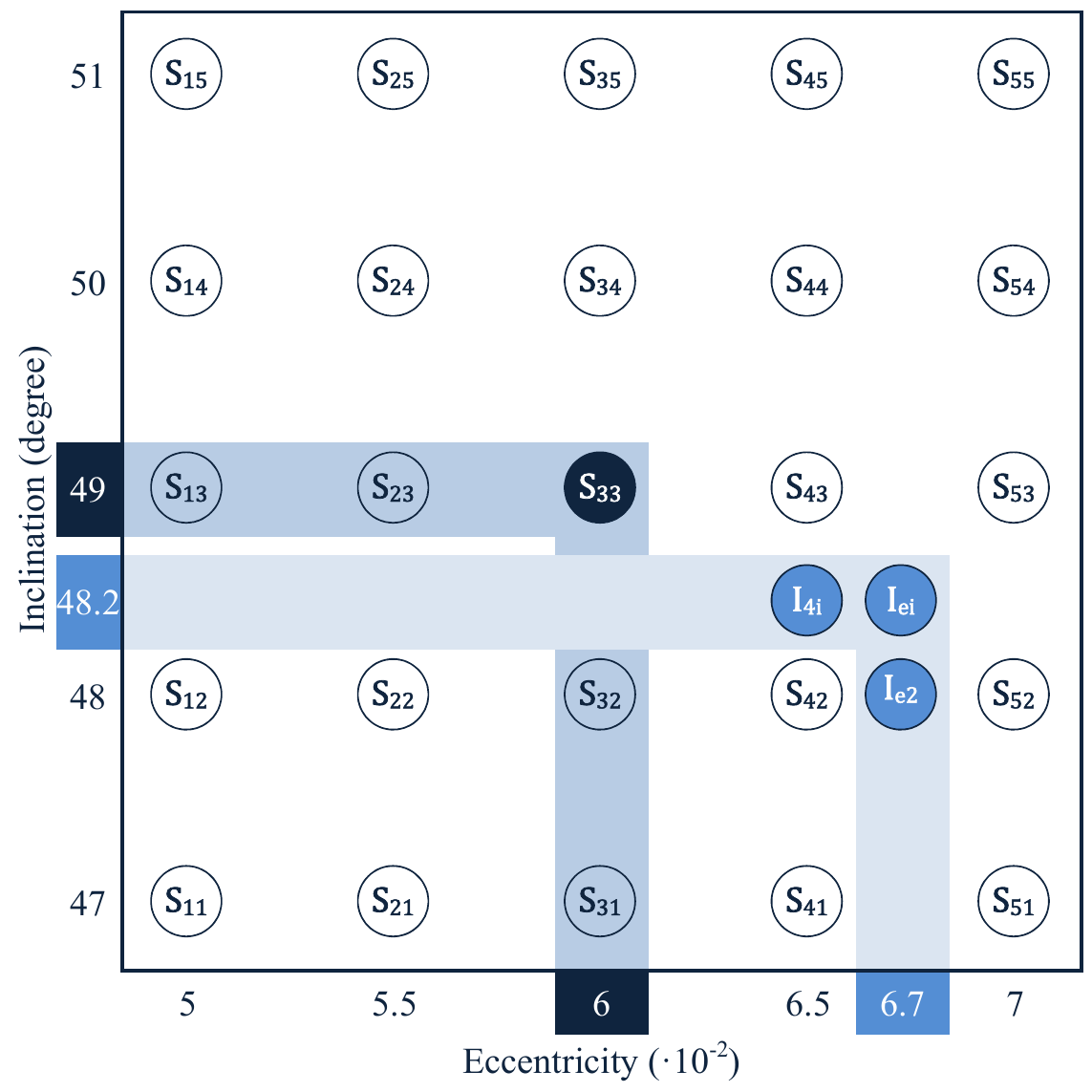}
 \caption{Grid of initial conditions $S_{ei}$ constructed around the studied satellite $S_{33}$, and intermediate initial conditions $I$.}
 \label{fgrid_S_G_I}
\end{figure}

After developing a hybrid propagator for the studied satellite, the effect of a slight change in the initial conditions will be analyzed. For that purpose, small variations in eccentricity and inclination will be considered. We construct a grid of initial conditions around the studied satellite, modifying its eccentricity in $0.5\times10^{-2}$ steps and its inclination in $1^\circ$ steps, as shown in Figure~\ref{fgrid_S_G_I}.

Next, we develop a new hybrid propagator for each initial condition $S_{ei}$ in the grid, following the steps described in the previous section for the studied satellite $S_{33}$. It is worth noting that control data are necessary for that process. Our final objective, in the next section, will be to verify the possibility to develop new hybrid propagators for initial conditions within the margins of the grid without having to follow the complete process, and hence with no need for control data.

\begin{figure}[!!htp]
\begin{subfigmatrix}{4}
\subfigure[ 1 day.]{\includegraphics{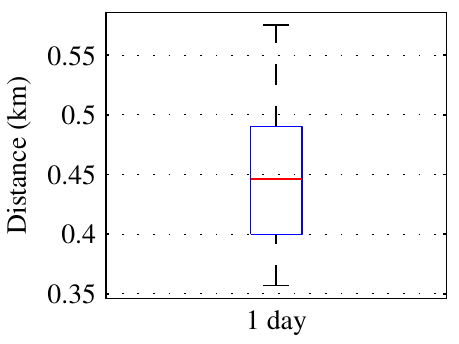}}
\subfigure[ 2 days.]{\includegraphics{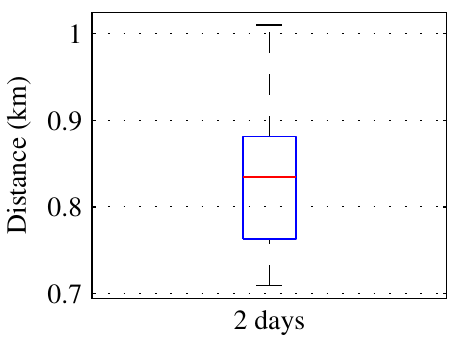}}
\subfigure[ 7 days.]{\includegraphics{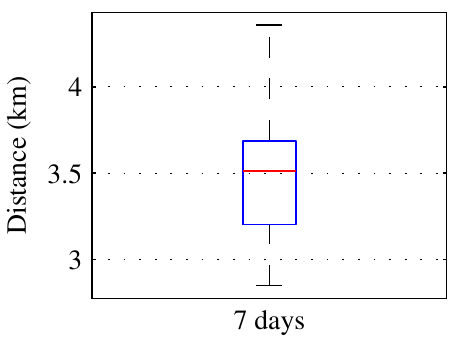}}
\subfigure[ 30 days.]{\includegraphics{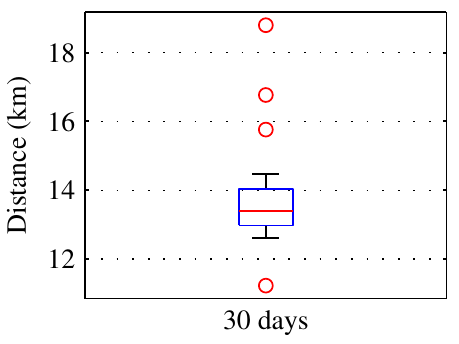}}
\end{subfigmatrix}
\caption{Boxplots of distance errors corresponding to the hybrid propagation of the grid initial conditions at different propagation horizons.}
\label{fboxplot_dist}
\end{figure}

We finish the creation of the grid hybrid propagators by analyzing their position errors with respect to the accurate numerical integration of their initial conditions. Figure~\ref{fboxplot_dist} shows their distribution after different propagation spans by means of boxplot graphics. It can be seen that their average values agree with those shown in Table~\ref{tdist_S} for the studied satellite $S_{33}$.

In general, the distributions of the position errors are symmetrical, showing little dispersion and only a few outliers in the case of a 30-day propagation horizon. All the initial conditions have similar dynamic behavior, which leads to the homogeneity of the obtained position errors. Such results constitute an appropriate scenario for the adaptation of the developed hybrid propagators to nearby initial conditions.

\section{Propagation of new orbits}

At this point, hybrid propagators for the studied satellite $S_{33}$ and its surrounding grid $S_{ei}$ have been developed. Now, we want to propagate nearby initial conditions $I$ which occupy intermediate positions within the limits of the grid (Figure~\ref{fgrid_S_G_I}).

The analytical theory in the first stage of the hybrid propagators is the same for all the cases. However, each set of initial conditions requires an individual Holt-Winters model in the forecasting stage of its hybrid propagator, aimed at modeling and predicting the effect of the $J_2$ perturbation under its particular conditions. In order to take advantage of the nearby hybrid propagators developed in advance, and also to avoid the need for control data, a new strategy is proposed: the interpolation of the parameters $A_T$, $B_T$, $S_{T-s+1},\ldots,S_{T-1},S_T$ of the intermediate $I$ Holt-Winters models from those corresponding to the studied satellite $S_{33}$ and its surrounding grid $S_{ei}$.

Several interpolation methods have been compared. Some of them only allow for one-dimensional interpolation, while others permit multi-dimensional interpolation. We perform comparisons on $I_{e2}$, which only needs one-dimensional interpolation because only one of its elements, the eccentricity, differs from the values in the grid.

In the first place, a \textit{weighted average} technique is used. We take the inverse of the difference in eccentricity as weight, and interpolate $I_{e2}$ Holt-Winters parameters from those of $S_{12}$, $S_{22}$, $S_{32}$, $S_{42}$, and $S_{52}$, which share the same inclination with $I_{e2}$. In the second place, the \textit{linear regression} method is applied, deducing $I_{e2}$ parameters from the nearest straight lines to $S_{12}$, $S_{22}$, $S_{32}$, $S_{42}$, and $S_{52}$ parameters. The third interpolation approach is performed through \textit{Lagrange polynomials}, by deducing $I_{e2}$ parameters from the fourth-order polynomials passing through $S_{12}$, $S_{22}$, $S_{32}$, $S_{42}$, and $S_{52}$ parameters. As it is known, the order of the Lagrange polynomials would increase if more initial conditions were available on the grid. Finally, \textit{spline interpolation} is used. This is the only considered method that permits multi-dimensional interpolation. The two-dimensional spline interpolation implemented in the Akima package \cite{aki2015_akima_r} of the R programming language \cite{r2015_sw_r}, which will be the method to be applied to the case of $I_{ei}$ because both its eccentricity and inclination differ from all the initial conditions present on the grid, is based on References~\cite{aki1978_irregfit} and \cite{aki1996_fit}.

\begin{table}
\caption{Position error (km) after the interpolated hybrid propagation of the intermediate initial conditions $I_{e2}$ through different interpolation methods.} \label{tdist_I11111_met}
\centering 
\begin{tabular}{ccccc}
\hline\noalign{\smallskip}
Propagation & Weighted & Linear & Lagrange & \multirow{2}{*}{Spline} \\
span & average & regression & polynomial & \\
\noalign{\smallskip}
\hline
\noalign{\smallskip}
$\phantom{l}1$ day &	$\phantom{14}9.070	$ & $\phantom{14}1.772$ & $\phantom{14}2.021$ & $\phantom{14}0.469$ \\[.0ex]
$\phantom{3}2$ days &	$\phantom{1}18.440	$ & $\phantom{14}3.563$ & $\phantom{14}4.070$ & $\phantom{14}0.836$ \\[.0ex]
$\phantom{3}7$ days &	$\phantom{1}66.222$ & $\phantom{1}12.460$ & $\phantom{1}14.237$ & $\phantom{14}3.498$ \\[.0ex]
30 days &	$272.777$ & $\phantom{1}48.653$ & $\phantom{1}55.715$ & $\phantom{1}13.598$ \\
\hline
\end{tabular}
\end{table}

Table~\ref{tdist_I11111_met} presents the results obtained for each of the four aforementioned interpolation methods by means of the position error of the interpolated hybrid propagators developed for $I_{e2}$. As can be seen, spline interpolation leads to the best results for all the propagation spans, followed by linear regression, Lagrange polynomial, and, finally, the weighted average technique, which yields the worst results.

The analysis of these interpolation methods applied to the other set of intermediate initial conditions that requires one-dimensional interpolation, $I_{4i}$, yields the same conclusions; therefore spline is selected as the interpolation method to be used. Then, an interpolated hybrid propagator is also developed for $I_{ei}$, making use of two-dimensional spline interpolation, as mentioned previously.

\begin{figure}[!!htp]
\begin{subfigmatrix}{4}
\subfigure[ 1 day.]{\includegraphics{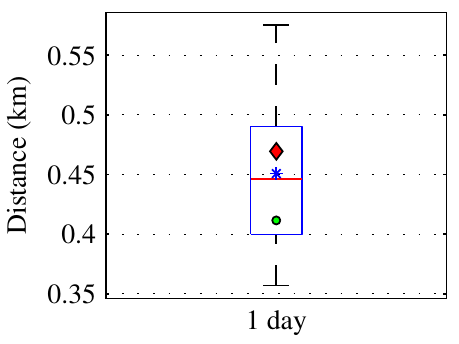}}
\subfigure[ 2 days.]{\includegraphics{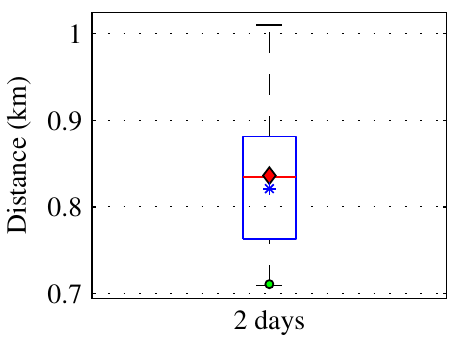}}
\subfigure[ 7 days.]{\includegraphics{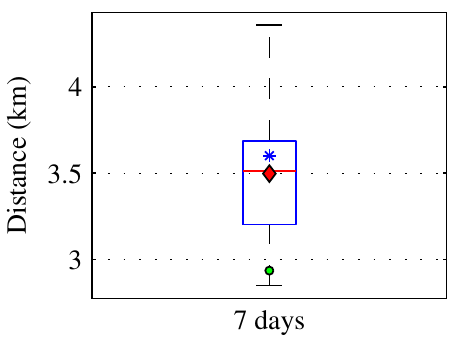}}
\subfigure[ 30 days.]{\includegraphics{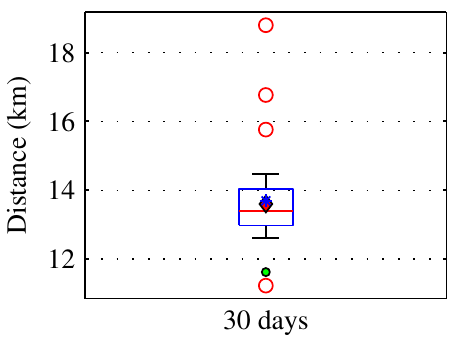}}
\end{subfigmatrix}
\caption{Position errors of the spline-interpolated hybrid propagation of the three intermediate initial conditions against the boxplot distributions of their corresponding grid position errors. Diamond, circle and star represent the initial conditions $I_{e2}$, $I_{ei}$, and $I_{4i}$, respectively.}
\label{fboxplot_dist_int}
\end{figure}

Figure~\ref{fboxplot_dist_int} represents the position errors obtained for the spline-interpolated hybrid propagation of the three intermediate initial conditions, and compares them with the distributions of the corresponding position errors for the hybrid propagation of the grid initial conditions (Figure~\ref{fboxplot_dist}). As could be expected, due to the homogeneous behavior of all the initial conditions in the grid, the position errors of the intermediate cases are very similar to the grid average. The case of the two-dimensional interpolation in eccentricity and inclination, $I_{ei}$, is remarkable because of its especially low errors, to the extent that it constitutes a low-error outlier for a propagation horizon of 30 days.

Tables~\ref{tdist_I11111}, \ref{tdist_I11113}, and \ref{tdist_I11112} compare the results of propagating the three sets of intermediate initial conditions $I_{e2}$, $I_{4i}$, and $I_{ei}$ through the mere analytic, the hybrid, and the spline-interpolated hybrid methods. In general, it can be verified that the latter propagators outperform the non-interpolated hybrid ones, especially in the case of the two-dimensionally spline-interpolated hybrid propagator for $I_{ei}$.

\begin{table}
\caption{Position error (km) after propagating the intermediate initial conditions $I_{e2}$.} \label{tdist_I11111}
\centering 
\begin{tabular}{cccc}
\hline\noalign{\smallskip}
Propagation & Analytic method & Hybrid method & Spline-interpolated hybrid  method \\
span & (Kepler) & (Kepler + $J_2$) & (Kepler + $J_2$)\\
\noalign{\smallskip}
\hline
\noalign{\smallskip}
$\phantom{l}1$ day &	$\phantom{1}1244.061$ &	$\phantom{1}0.600$ &	$\phantom{1}0.469$ \\[.0ex]
$\phantom{3}2$ days &	$\phantom{1}2472.668$ &	$\phantom{1}0.840$ &	$\phantom{1}0.836$ \\[.0ex]
$\phantom{3}7$ days &	$\phantom{1}8165.104$ &	$\phantom{1}3.711$ &	$\phantom{1}3.498$ \\[.0ex]
30 days &	$14504.581$ &  $14.982$ & $13.598$ \\
\hline
\end{tabular}
\end{table}

\begin{table}
\caption{Position error (km) after propagating the intermediate initial conditions $I_{4i}$.} \label{tdist_I11113}
\centering 
\begin{tabular}{cccc}
\hline\noalign{\smallskip}
Propagation & Analytic method & Hybrid method & Spline-interpolated hybrid  method \\
span & (Kepler) & (Kepler + $J_2$) & (Kepler + $J_2$) \\
\noalign{\smallskip}
\hline
\noalign{\smallskip}
$\phantom{l}1$ day &	$\phantom{1}1232.193$ &	$\phantom{1}0.533$ &	$\phantom{1}0.451$ \\[.0ex]
$\phantom{3}2$ days &	$\phantom{1}2449.259$ &	$\phantom{1}0.819$ &	$\phantom{1}0.821$ \\[.0ex]
$\phantom{3}7$ days &	$\phantom{1}8098.784$ &	$\phantom{1}3.828$ &	$\phantom{1}3.601$ \\[.0ex]
30 days &	$14510.092$ &	$15.090$ &	$13.691$ \\
\hline
\end{tabular}
\end{table}

\begin{table}
\caption{Position error (km) after propagating the intermediate initial conditions $I_{ei}$.} \label{tdist_I11112}
\centering 
\begin{tabular}{cccc}
\hline\noalign{\smallskip}
Propagation & Analytic method & Hybrid method & Spline-interpolated hybrid  method \\
span & (Kepler) & (Kepler + $J_2$) & (Kepler + $J_2$) \\
\noalign{\smallskip}
\hline
\noalign{\smallskip}
$\phantom{l}1$ day &	$\phantom{1}1240.220$ &	$\phantom{1}0.561$ &	$\phantom{1}0.411$ \\[.0ex]
$\phantom{3}2$ days &	$\phantom{1}2465.050$ &	$\phantom{1}0.823$ &	$\phantom{1}0.711$ \\[.0ex]
$\phantom{3}7$ days &	$\phantom{1}8143.626$ &	$\phantom{1}3.724$ &	$\phantom{1}2.936$ \\[.0ex]
30 days &	$14506.802$ &  $14.950$ &  $11.619$\\
\hline
\end{tabular}
\end{table}

\section{Conclusion and future work}

In this work, we have presented an advance in the hybrid propagation methodology. Hybrid propagators are composed of an integration theory plus a forecasting technique. The latter is developed from control data so as to complement the approximation generated by the former by modeling and reproducing the missing dynamics. We have explored the possibility of deducing the forecasting stage directly from other hybrid propagators developed for surrounding initial conditions. This approach avoids the need for control data, and makes it possible to have a grid of hybrid propagators prepared in advance for a region of initial conditions of interest. We have verified that the spline interpolation of the parameters of an additive Holt-Winters forecasting method from nearby hybrid propagators yields similar accuracy, or even better, to a non-interpolated hybrid propagator. The study has been conducted using the \textit{main problem} of the artificial satellite theory as the propagation model, with the forecasting stage modeling the complete $J_2$ effect.

At present, we are testing the hybrid propagation methodology considering neural networks instead of the Holt-Winters algorithm as time series forecasters. During the second semester of 2017, a competition organized by the European Space Agency will be launched through the Advanced Concepts Team competition website, \textsc{Kelvins},\footnote{\url{https://kelvins.esa.int/}} in order to encourage the machine learning community to get involved and participate in the problem.

\section*{Acknowledgments}

This work has been funded by the Spanish State Research Agency and the European Regional Development Fund under Project ESP2016-76585-R (AEI/ERDF, EU). Support from the European Space Agency through Project Ariadna Hybrid Propagation (ESA Contract No. 4000118548/16/NL/LF/as) is also acknowledged.

\bibliographystyle{splncs03}
\bibliography{bibliography}

\end{document}